
\baselineskip =20pt
\font \big = cmbx12 at 20pt
{\big \centerline {Lorentz-invariant Bohmian Mechanics}}
\vskip 1cm
\centerline                     {\it Euan J. Squires}
\centerline {\it Department of Mathematical Sciences}
\centerline {\it University of Durham}
\centerline {\it Durham City}
\centerline {\it England, DH1 3LE}
\vskip 1.5cm
A derivation of the Bohm model, and some general comments about it,
 are given. A modification of the model which is formally local and
 Lorentz-invariant is introduced, and its properties studied for a
 simple experiment.
 \vskip 1cm
\noindent 1. {\bf The Bohm model as a simple realistic quantum theory.}

\vskip .7cm

Non-relativistic quantum
 mechanics, as it is presented in almost all text-books, is a theory
which is either incorrect or incomplete, even within the domain where
non-relativistic approximations are adequate. It apparently gives the
correct predictions for the outcome of measurements, but nowhere within it
does it contain those outcomes, or allow any description of the processes
whereby they are produced. By far the simplest method of curing this
problem is that introduced by David Bohm in 1952. Here  quantum
theory is correct, but it is incomplete. The complete theory has, in
addition to the wavefunction, trajectories for individual particles
exactly as in classical mechanics. Indeed, as we discuss in the next
 section, it is possible to regard the
Bohm model as being an extension of classical mechanics.

We shall first consider the simple
 logical steps which allow us to derive the Bohm model directly from the
rules of orthodox quantum theory. The starting point
here is to note that all observations are in reality observations of
{\it position}. We deduce the results of measurements of other
quantities by an observation of position (consider for example the
measurement of a spin projection by a Stern-Gerlach device). Exactly {\it
why} this is so is an interesting question itself (Squires, 1990).
Next we recall that quantum theory gives statistical predictions. Thus we
require a model in which objects at all times have positions, and which
 gives the correct statistical distribution of these
positions.

To see what this means we suppose that we have N particles with positions
${\bf x}_i(t)$, where $t$ is the time. We can represent these by the vector
 ${\bf X(t)}$ in the $3N$ dimensional configuration
space.
If we denote the probability distribution of the positions at time $t$ by
 $\rho ({\bf
X}, t)$, then the condition that $\rho ({\bf X}, t +dt)$
 gives the probability distribution
at $t+dt $ is clearly  $$\rho ({\bf X}, t)d^{3N}{\bf X}
 = \rho ({\bf X} + { \bf \dot
 X}dt, t + dt)d^{3N}{\bf X}(t + dt),\eqno (1.1)$$
 which leads to the standard continuity equation
$$\nabla .(\rho {\bf \dot X}) \equiv \sum _i \nabla _i .(\rho {\bf \dot
x_i})= - {\partial \rho \over \partial t},\eqno
(1.2)$$where ${\bf \dot X}({\bf X}, t)$ is the vector field giving the
velocity of a particle at ${\bf X}$.

The rhs of eq.(1.2) can be evaluated if we write the density in terms
of a wavefunction evolving according to the Schr\"odinger equation:$$\rho
= \Psi ^*\Psi,\eqno (1.3)$$
 with $$i\hbar \dot \Psi = \left (-{\hbar ^2 \over 2}\sum
_i {1 \over
m_i}
{\partial ^2 \over \partial x^2_i} + V \right )\Psi. \eqno (1.4)
 $$A simple calculation yields $$\eqalignno
 {{\partial \rho \over \partial t}&=
{i\hbar \over 2}\sum _i {1 \over m_i}\nabla _i.\left [\Psi ^* \nabla
_i\Psi -\Psi \nabla _i\Psi ^*\right]&(1.5)\cr&=-\sum _i\nabla _i.{1 \over
m_i}\Re\left (\Psi ^* {\bf p}_i\Psi \right ),&(1.6)\cr}$$where
${\bf p}_i$ is the momentum operator for the $i^{th}$
particle, $${\bf p}_i = -i\hbar \nabla _i.\eqno (1.7)$$

 Comparing eq. (1.6) with (1.2)
we see that  $$\dot {\bf x}_i
 = {1 \over m_i}\Re \left ({\bf p}_i\Psi \over \Psi
\right) + ({1 \over \rho })
 {\bf c}_i, \eqno (1.8)
$$where the  ${\bf c}_i$ are arbitrary vectors which satisfy $$\sum _i \nabla
_i.{\bf c}_i = 0. \eqno (1.9)$$

To obtain the Bohm expression we take the simplest form, i.e.,
zero, for the ${\bf c}_i$. This can be justified essentially on the grounds
of simplicity, together with the fact that any arbitrary vectors ${\bf c}$,
not dependent on $\Psi $, would break rotational invariance, and would
give the very unphysical result that the velocity would go to infinity as
the density went to zero. It is also the obvious choice if we use the
standard form for the probability current, which is similarly
underdetermined.

 Before proceeding we note, however, that there is a simple case where
the neglect of the final term is rather less
 ``natural''. This is when $\Psi $,
and hence $\rho $, is independent of time. Then the rhs of eq. 1.2
is zero, which
would suggest zero velocity as the natural solution. The fact that the
Bohm model need not give zero velocity in such a situation may be
significant in quantum cosmology (Valentini, 1992, Vink, 1992: Squires,
1992, 1994). Here, according to the Wheeler-deWitt
 equation, the wavefunction of the universe (which is the only
 wavefunction that actually {\it exists}!) is independent of time. This
 is a consequence of the fact that the theory must be invariant under
 reparameterisation of time. For any real solution of this equation,
 the straightforward
 generalisation of Bohmian mechanics to quantum cosmology
 predicts zero velocities, i.e., a universe in which
 nothing ever moves. Presumably this is not a good prediction! There is of
 course an analogous prediction in the microscopic world where for
 example the model predicts that an electron in the ground state of a
 hydrogen atom does not move. In this case, however, the result is
 not a problem because we know that predictions for the results of {\it
 measurements} of the electron velocity, which will be related to
 positions of certain probes, will be correct. There is no similar
 escape in  the cosmological case - a stationary universe
 is simply a stationary universe! Thus it is essential to select a
 (non-trivially) complex solution of the Wheeler-deWitt equation, and to
 use the fact
 that such a wavefunction can give non-zero velocities, even if
 the wavefunction itself is constant.

Eqs. (1.4)  and (1.8), with the ${\bf c}_i$ equal to zero,
completely define the theory. Provided that in any experiment  the initial
distribution of positions agrees with that given by the quantum rule at
the initial time then they will do so at the end of the experiment, which
then guarantees that the model will always agree exactly with the
predictions of quantum theory. Note, especially, that the model
automatically avoids the hidden-variable ``no-go'' theorems (Bell, 1966,
Hardy, 1995,  Clifton and Pagonis, 1995); in other
words, the form
of eq.(1.8) ensures that there is the necessary contextuality of
measurements (see remarks below eq. 3.2).

To find an analogue of Newton's second law of motion we put
 $$\eqalignno
  {{\bf \ddot x}_i&= {\partial {\bf \dot x}_i \over \partial t} +
\sum _j {d{\bf x}_j \over
  dt}.\nabla _j{\bf \dot x}_i&(1.10)\cr
  &={1 \over m_i}\Re \left[-i\hbar {\partial \over \partial t} \left (
  {\nabla _i \Psi \over \Psi} \right)  + \sum _j{\bf \dot x}_j
   .\nabla _j  \left ({-i\hbar \nabla _i \Psi
   \over \Psi } \right) \right ],&(1.11)\cr}$$where we have used
  eqs. (1.7) and (1.8). After a little
   rearrangement this becomes $$m_i{\bf \ddot x}_i = -\nabla _i \left [V
   +Q\right ],\eqno (1.12)$$which is Newton's second law with
    the potential, $V+Q$, given by $$V + Q=
   \Re \left ({H \Psi \over \Psi } \right) -\left (1 \over 2 \right)
   \sum _j m_j
   {\bf \dot x}_j^2,\eqno (1.13)$$ where $H$ is the hamiltonian. We can
   separate out the ``quantum potential'', $Q$, by using eq. 1.4. This
   gives $$Q = \Re \sum _j {1 \over 2m_j} \left ({{\bf p}_j^2 \Psi \over
    \Psi } -
    m_j ^2
   {\bf \dot x}_j^2\right ),\eqno (1.14)$$

This equation reveals some interesting features of the
model. For example, if we replace the operator ${\bf p}$ by its classical
 value $m {\bf \dot x}$ the quantum potential becomes zero, so in this
 sense the Bohm equation for the velocity may be said to contain
 Netwon's second law. Nevertheless, although we expect that the quantum
 potential should be a  small quantum correction, it
 clearly exactly cancels (up to an irrelevant constant) the
``classical'' potential in the case when the state is an energy
eigenstate with a constant phase, as occurs in particular for a lowest
 energy bound state.

 It is important to note, however,
that this step of introducing the potential
is not necessary. Unlike Newtonian mechanics, Bohmian
mechanics gives an  equation for the velocities, not the accelerations.
The initial conditions for a Bohmian universe are not positions and
velocites, but positions (together of course with the wavefunction). Even
the positions are not free, but have to satisfy the constraint that, at
some initial time, they will give probability distributions consistent
with the Born rule (see D\"urr, Goldstein and Zhangi, 1992,  for a detailed
discussion).

\vskip 1cm
\noindent {\bf 2. Comparison with classical mechanics.}
\vskip .7cm

Before proceeding it is of
 interest to see how the Bohm model relates to other models.
  In my opinion the model looks much more convincing if we
 emphasise its similarity to {\it classical}
  mechanics rather than to (orthodox) {\it quantum}
 mechanics. The ontology of the model is that of classical mechanics; it
 has real particles, which at all times have positions. The law
 describing how the particles
 move has the same form as in classical mechanics; the trajectories
 are defined by  Newton's second
 law of motion.

On the other hand, there is none of the indeterminism, or special role
of observations, which are characteristic of quantum theory.

There are, of course, differences between Bohmian mechanics and
Newtonian mechanics, although to some extent these can be regarded as
additions to the latter rather than changes. One such difference is that,
as we have noted, the second order equation of motion can be integrated
to give a first order equation, without any need for additional boundary
conditions. To appreciate the significance of this point we might imagine
 a world in which time is discrete. Then it is clear that in Newtonian
 mechanics, the equations plus conditions at one time do not completely
 determine the state at future times. In the Bohm model however they do.

 The other difference is that there is an additional
``quantum'' force. Of course classical physics is easily able to cope
with new forces, but this particular addition is not as
 innocent as it sounds. The quantum force is unlike all the other forces
  in nature
because it is not derivable from the positions of the particles. We can
compare it with, for example, gravity. The
force of gravity acting on a given particle is a unique function of the
positions of the other particles, which act as sources in the Poisson
equation. This of course is only true if we neglect the ``complimentary
function'' (i.e. solution of the Laplace equation), a procedure
 which we normally
justify by imposing some sort of boundary condition at infinity.
The quantum potential, however,
 is derived from the Schr\"odinger equation in
which there are no sources, so the analogue of the neglected term is
here everything. Also the actual quantum force is independent of the
{\it magnitude} of the field from which it derived (see eq. 1.14).

These differences strongly suggest that there is some underlying theory,
which is not quantum theory, and not classical mechanics, but which
combines (and explains?) certain aspects of both. It is one of the great
merits of the Bohm model  that, in addition to  giving a proper, respectable,
explanation of all quantum phenomena, it encourages speculation
about such an underlying theory, and even about possible
 theories which give results different to those of standard quantum
 theory. An example is discussed below.

 \vskip 1cm

\noindent {\bf 3. Relativistic invariance and the Bohm model.}

\vskip .7cm

The Bohm model exposes the non-locality which has long been recognised as
 one of the significant features of the difference between
  quantum mechanics and classical physics.
   Bell's theorem demonstrated that this non-locality is
 not peculiar to the particular form of realistic model used by Bohm.
 Any ``completion'' of quantum theory which is consistent with all its
 predictions must be non-local. Since the experimental tests
  (e.g., Aspect, et al., 1981, 1982a,b)  seem to
   agree with quantum theory, most physicists  have come to
 accept this non-locality in some form or other. Although the
 discussions are normally carried out in a non-relativistic context, it
 is clear that agreement with the predictions of quantum theory strongly
 suggests that a realistic model should be non-Lorentz invariant, in
 particular should require that there exists a preferred frame (Hardy,
 1992, Hardy and Squires, 1992).

The need for such a frame is  of course evident in the Bohm model
because the expression for the velocity of one particle (eq. 1.8)
requires
knowledge of the position of all the others {\it at the same time}, which
clearly is not a frame-independent concept. However, the fact that the
model so clearly reveals the non-locality means that it also shows how
it might be removed, at the cost of course of a failure to agree at all
times with the predictions of quantum theory. The idea is suggested by
the analogy with a classical potential given in section 2. Any classical
potential, e.g., the electrostatic potential, is also defined in the
configuration space of the particles, and requires simultaneous
positions. However, we know how this is dealt with in a proper
relativistic treatment: we use the ``retarded'' potential in which the
positions are determined on the backward light cone.

It is possible to do something similar in the Bohm model (Squires, 1993,
Mackman and Squires, 1995)
 and to replace
eq. 1.8 by $$\dot {\bf x}_i(t_i)
 = {1 \over m_i}\Re \left ({\bf p}_i\Psi ({\bf x}_1(t_1), {\bf
 x}_2(t_2)....) \over \Psi ({\bf x}_1(t_1), {\bf
 x}_2(t_2)....)
\right),\eqno (3.1)$$where $$t_k = t_i - {|{\bf x}_i(t_i) - {\bf
x}_k(t_k)| \over c}.\eqno (3.2)$$

In equation (3.1) we have ignored the explicit time
dependence  of the wavefunction, and it is not clear how we should treat
this. Part of the problem is that we are working within the framework of
non-relativistic quantum theory. At the
 fundamental level, we could perhaps take refuge in the fact noted
above that the actual wavefunction of the universe is constant. Further
work is needed here but for the moment we shall ignore the problem. In
the example discussed below an unambiguous procedure suggests itself.

It is clear that eq. (3.1)
 goes some way towards removing the obvious non-locality from the Bohm
model, and it is  important to study the nature of its inevitable
disagreement with quantum
theory predictions. In principle this is possible
 because the model allows explicit
calculations.

 Consider, for example, an experiment in one space
dimension in which a photon is emitted from an origin
in the form of two wave-packets, of equal size,
one travelling in the positive direction and one in the negative. We
suppose that there are photon detectors at positions $l$ and $-l$, and
that
the purpose of the experiment is to determine in which direction
 the photon is observed to travel, i.e.,  which detector actually
``sees'' the photon. The difficulty with this type of discussion is how
we model a real photon detector. One simple
possibility (see Squires, 1993, and
Squires and Mackman, 1994) is to take the detectors to be free
particles, initially in zero-momentum gaussian wave-packets, which receive
a momentum $p$ when they detect a photon of momentum $p$. Thus, just
after the photon is emitted, the wavefunction of the system is given by
$$|\Psi > = 2^{-{1 \over 2}}[\phi _L(y)  +
\phi _R(y)] \psi _L(x_L)\psi _R(x_R), \eqno (3.3)$$
where $\phi _{L,R}(y)$ is the part of the photon wavefunction moving
 to the $L,R$ respectively, and $\psi _{L,R}$ are the two detector
 states given by $$\psi _{L,R}(x_{L,R}) =({a \over \pi })^{{1 \over
 4}}\exp [-{1 \over 2}a(x_{L,R} \pm l)^2].\eqno (3.4)$$

At a later time, after the photon wave-packets have interacted with the
detectors, the wavefunction has the form: $$|\Psi > = 2^{-{1 \over 2}}[\phi
 _L(y) \psi ^{-p}_L(x_L)\psi _R(x_R) +
\phi _R(y) \psi _L(x_L)\psi ^p_R(x_R)], \eqno (3.5)$$
where  the $\psi ^{\pm p}$ now represent moving wavepackets, e.g.,
$$\psi ^{-p} (x_L) = ({a \over \pi })^{{1 \over 4}} \exp [i(x_Lp +{p^2t
\over 2m}) - {1 \over 2}a(x_L + l +{pt \over m})^2]. \eqno (3.6)$$Note
that in the last expression we have neglected
the quantum evolution of the free
detector state.

Now we recall that in a spin measurement, as for example in the
 experiments of Aspect et al., the actual outcome, i.e., the value of
 the spin that is observed, is determined in the Bohm model by the
 value(s) of the,  so-called, hidden variable(s) in the detector. To
 study something analogous to this we
 suppose that
there are no photon trajectories (this is the case in at least some
versions of the Bohm model). Then, as we shall see below, the
measurement outcome, which we refer to as the position of the photon,
again depends upon the values of
the hidden variables of the detectors, i.e., the positions of the
particles.

 First, it is necessary to modify the standard Bohm
formula for particle velocities by integrating over the positions of those
 particles without trajectories (Bell, 1981, Squires and Mackman, 1994).
  Thus, in general,
  eq. (1.8) must be replaced by
 $$\dot {\bf x}_i
 =\Re \left ( {\int d^3{\bf y}\Psi ^* {\bf p}_i\Psi \over m_i\int d^3
 {\bf y}\Psi
 ^*\Psi }\right)
 , \eqno (3.7)
$$In our experiment
 this leads to the result $$m\dot x_L = \Re \left({|\psi
_R|^2\psi ^{-p*}_Lp^{op}_L\psi _L^{-p} + |\psi
_R^p|^2\psi ^*_Lp^{op}_L\psi _L \over |\psi _R|^2|\psi _L^{-p}|^2 +
|\psi _R^p|^2|\psi _L|^2}\right )
\eqno (3.8)$$ and a similiar equation for  $\dot
x_R$, where $m$ is the mass of the detector particle.
 Here we have assumed that there is no overlap between the right
and left moving photon wave-packets.  This will be approximattely true
for any reasonable definition of the space-time surface implicit in eq.
(3.7).

We first use the non-retarded, and hence  non-local, Bohm model. Then,
inserting the previous wavefunctions into Eq. (3.8), we find
 $$\dot u = \left (1 + \exp [2t(v-u)]\right
)^{-1}, \eqno (3.9)$$and $$\dot v= \left (1+\exp[2t(u-v)]\right
)^{-1},\eqno (3.10)$$where we have simplified the notation by using
units in which $a= {p \over m} = 1$ and by defining $$u=x_R-l\eqno
(3.11)$$and $$v=-(x_L+l).\eqno (3.12)$$

Adding (3.9) and (3.10) we obtain $\dot u + \dot v = 1$, hence
 $$u + v = t + u_0 + v_0, \eqno (3.13)$$where $u_0$ and $v_0$ are the
 values of $u$ and $v$ at the time when the interaction occurs, taken to
 be $t=0$. If we substitute (3.13) into (3.9) and (3.10) we find $$\dot u
  = \left (1 + \exp [-2t(2u-t-u_0-v_0)]\right
)^{-1}, \eqno (3.14)$$and $$\dot v= \left (1+\exp[-2t(2v-t-u_0-v_0)]\right
)^{-1}.\eqno (3.15)$$

We compare these results with what we obtain with only one detector, say
the one at the left. Then the Bohm equation would give$$\dot v = \left
(1 + \exp [-t(2v-t)]\right )^{-1}.\eqno (3.16)$$
 Clearly the small $t$ behaviour is
$v\simeq {1 \over 2}t +v_0$, leading to $$\dot v \simeq \left ( 1 + \exp
[-2tv_0]\right )^{-1},\eqno (3.17)$$ for small $t$. It follows that this
detector will record the photon (in the sense that the detector particle
will continue to move, with velocity approaching $1$),
 if $v_0 > 0$, but not if $v_0 < 0$. Incidently
we can here see the fact that an initial distribution agreeing with
quantum theory will give the correct quantum theoretic outcome:
 such an initial
distribution will have the $v_0$ equally distributed between positive and
negative values, leading to half the photons being detected in the left
detector, as required.

If we treat eq. (3.15) in a similar way we find $$\dot v \simeq \left ( 1 +
\exp
[-2t(v_0 - u_0)]\right )^{-1}.\eqno (3.18)$$Hence in this case the
condition that the left detector records (fails to record) the photon
is that $v_0 - u_0$ is positive (negative). Clearly the opposite is true
for the right detector, so (as required) one, and only one,
 detector will see the
photon. Again an initial distribution agreeing with quantum theory will
have $v_0 - u_0$ positive and negative with equal frequency, so giving
the expected ouput results. We note also that this example reveals the
contextuality of this version of
the Bohm model: comparison of (3.17) with (3.18) shows
that the result emerging from the right
detector, say, is affected by the presence of the left detector.

The differential equations (3.14) and (3.15) can in fact be solved
directly to give, for example, $$\int _{-(u_0 - v_0)
 \over 2}^{(u_0-v_0) \over 2}dye^{-2y^2}=\int _{t-(u-{(u_0+v_0)\over
 2})}^{u-{(u_0 +v_0) \over 2}}dye^{-2y^2}.\eqno (3.19)$$ Clearly, if
 $(u_0 - v_0)$ is positive, then $(u - t)$ must remain constant as $t$
 becomes large; on the other hand if it is negative, then $u$ becomes
 constant and small for large $t$. This confirms that we will obtain the
 expected measurement outcomes.

We must now consider what happens in this experiment if we use the
retarded Bohm model. Since we have well-localised wave-packets we can solve
the problem noted below eq. 3.1 by evaluating the wave-packet from eq.
3.6  at
the appropriate retarded time as well as retarded position.
 We define $T$, the time for light to travel from
one detector to the other, according to $$T = {2l \over c}. \eqno
(3.20)$$Then, for $t<T$, the two detectors will behave as if the other
one was not present. Hence, for $t < T$, $$\dot u(t)=\left (1 + \exp
[-t(2u(t)-t)]\right )^{-1}\eqno (3.21)$$ and $$\dot v(t)=\left (1 + \exp
[-t(2v(t)-t)]\right )^{-1},\eqno (3.22)$$where, for reasons which will
be immediately evident, we have explicitly
written the time arguments.

Thus, up to $t=T$, the detectors behave independently and record the
presence, or otherwise, of the photon strictly according to their own
initial position. Clearly this means that in some cases ``wrong''
results are occuring, i.e., both, or neither, detector is seeing the
photon. However, at $t=T$, the situation changes because information
 about the presence of the other detector becomes
available. Thus, for $t>T$, $$
\dot u(t)= \left (1 +  \exp [-2u(t)t + 2v(t-T).(t-T) + T.(2t - T)] \right
)^{-1}\eqno (3.23)$$and  $$
\dot v(t)= \left (1 +  \exp [-2v(t)t + 2u(t-T).(t-T) + T.(2t - T)] \right
)^{-1}.\eqno (3.24)$$ Note that, as expected, these equations agree with
eqs. (3.9) and (3.10) if $T$ is put equal to zero.

I have not been able to solve these equations analytically but it
 is clear that
they give the expected results (Squires, 1993). In particular,
 if $T$ is sufficiently small, and $|v_0 - u_0|$ sufficiently large,
 then again, one, and only one, detector will record the photon. There
 are, however, circumstances in which both, or neither, detector will
 see the photon., This would correspond to a ``wrong'' result, in the
 sense that the predictions of orthodox quantum theory would be
 violated.

More precisely, the condition that there will be a significant number of
``wrong'' results (zero or two photons), is that $$T \geq |v_0 -
u_0|_{typical}. \eqno (3.25)$$With the units restored this means$${l
\over c} \geq {m \over p a^{{1 \over 2}}}.\eqno (3.26)$$If we now assume
 that the detector acquires all the initial momentum of the detected
 photon then $p={\hbar \over \lambda}$ so the condition for wrong
 results becomes $${l\hbar \over mcd\lambda }\geq 1,\eqno (3.27)$$
 where $d\sim a^{-1/2}$ is the
 spatial spread of the initial wavefunction of the detector particle.
In fact numerical solutions of eqs. (3.23) and (3.24) show that when
$${l\hbar \over mcd\lambda }= 1,\eqno (3.28)$$
then about one
 in ten events give wrong results (H.Movahhedian, private communication).

In a typical experiment the separation $l$ is only a few metres, so if
for $m$ we take the mass of a macroscopic pointer, it is clear that the
condition in eq. (3.27) is not satisfied. On the other hand if we suppose the
detection comes about by the photon being absorbed by an electron, then
for an optical photon, the LHS of (3.27) is of the order of ${10^{-5}
\over d}$, with $d$ measured in metres. If the electron is initially
confined to within an atomic distance then clearly this is much greater
than 1, so there will be many events in which both, or neither,
electron records the photon. Of course, we would certainly not directly
observe a single electron, and it would be essential here to use a
device which was genuinely responsive to the electron trajectory (this is
not always a trivial issue - see, for example, Englert, et al., 1992 and
Dewdney, et al., 1993).

The tentative conclusion of this analysis is that it is unlikely that
deviations from the quantum theory results, which would arise in our
retarded model, would have been seen in any experiments that have been
performed. Nevertheless a better analysis of the actual experiments is
required, and such an analysis could well reveal the possibility of
realistic tests for retarded effects in future, carefully designed
experiments. Such experiments would need the largest possible values of
$l$, detectors where the effective ``mass'' of the detector is as small
as possible, and of course efficient detectors.
\vfil \break
     \vskip 1.5cm
\noindent {\bf REFERENCES}
               \vskip .7cm

         \noindent
Aspect, A., Grangier, P. and Roger, G. (1981) ``Experimental tests of
realistic theories via Bell's theorem'', {\it Physical  Review Letters}, {\bf
47}, 460-463.  \hfil \break
\noindent Aspect, A., Grangier, P. and Roger, G. (1982a) ``Experimental
realisation of Einstein-Podolsky-Rosen-Bohm {\it gedankenexperiment}:
a new violation of Bell's inequalities'',
 {\it Physical  Review Letters}, {\bf
49}, 91 -94.  \hfil \break
\noindent Aspect, A., Dalibard, J. and and Roger, G.  (1982b)
``Experimental test of Bell's inequalities using time-varying
detectors'', {\it Physical  Review
Letters}, {\bf 49}, 1804-1807.\hfil\break
\noindent Bell, J.S. (1966) ``On the problem of hidden variables in
quantum theory'', {\it Reviews of Modern Physics}, {\bf 38},
447-52.\hfil\break
\noindent Bell, J.S. (1981) in {\it Quantum Gravity 2}, ed. Isham C., Penrose,
R. and Sciama, D. (Clarendon Press, Oxford) p. 611. \hfil \break
\noindent D\" urr, D., Goldstein, S. and Zhangi, N. (1992) ``Quantum
equilibrium and the origin of absolute uncertainty'', {\it Journal of
Statistical Physics}, {\bf 67}, 843-907.\hfil \break
\noindent Dewdney, C., Hardy, L. and Squires, E.J. (1993) ``How late
measurements of quantum trajectories can fool a detector'', {\it Physics
Letters}, {\bf A184}, 6-11.\hfil \break
\noindent Englert, B., Scully, M.O., S\"ussmann, G. and Walther, H. (1992)
``Surrealistic Bohm  trajectories'', {\it Zeitschrifft f\"ur
Naturforschung }, {\bf 47a}, 1175-86. \hfil \break
\noindent Hardy, L. (1992) ``Quantum mechanics, local realistic
theories, and Lorentz-invariant realistic theories'', {\it Physical
Review
Letters,} {\bf 68}, 2981-2984. \hfil
\break
\noindent Hardy, L. (1995) this volume.....\hfil \break
\noindent Hardy L. and Squires, E.J. (1992) ``Hidden variable theories
violate Lorentz invariance'', {\it Physics Letters}, {\bf A168}, 169-173.
 \hfil \break
 \noindent Mackman, S. and Squires, E.J. (1995) ``Lorentz-invariance and the
retarded Bohm model'', {\it Foundations of Physics},  {\bf 25}, 391-397.
 \hfil \break
\noindent  Pagonis, C. and Clifton, R. (1995) ``Unremarkable
contextualism: dispositions in the Bohm theory'',{\it Foundations  of
Physics,}
{\bf 25}, 281-296. \hfil \break
\noindent Squires, E.J. (1990) ``Why is position special?'',
 {\it Foundations of Physics Letters}, {\bf 3}, 87-93.\hfil \break
\noindent Squires, E.J. (1992) ``An apparant conflict between the
deBroglie-Bohm
model and orthodoxy in quantum cosmology'', {\it  Foundations of
Physics Letters}, {\bf 5}, 71-75.\hfil \break
\noindent Squires, E.J. (1993) ``A local hidden-variable theory that
FAPP agrees with quantum theory'',
 {\it Physics Letters}, {\bf A178}, 22-26.\hfil \break
\noindent Squires, E.J. (1995) in {\it Fundamental Problems in Quantum
 Theory}, ed. Greenberger, D.M. and Zeilinger, A., {\it Annals of the
 New York Academy of Sciences} {\bf 755}, 451-465.\hfil \break
\noindent Squires, E.J. and Mackman (1994) ``The Bohm model with
fermion-boson correlations'', {\it Physics Letters,}{\bf A185}, 1-4.\hfil
\break
 \noindent Valentini, A. (1992)  ``On the pilot-wave theory of classical,
 quantum and sub quantum physics'', Thesis submitted for the degree of
 ``Doctor Philosophiae'', SISSA, Trieste, preprint. \hfil
 \break
 \noindent Vink, J.C. (1992) ``Quantum potential interpretation of the
 wavefunction of the universe'', {\it Nuclear  Physics},
  {\bf B369}, 707-728. \hfil
 \break

 \end